\documentclass[prl,twocolumn,superscriptaddress,showpacs,floatfix,shortbibliography]{revtex4-1}
\usepackage{graphicx}
\usepackage{bm}
\usepackage{color,soul}
\usepackage{amsmath,amssymb,amsfonts,float,graphics,epsfig,epstopdf,color,verbatim,tabularx,bm,multirow,appendix,hyperref}
\begin{document}
% Use the \preprint command to place your local institutional report
% number in the upper righthand corner of the title page in preprint mode.
% Multiple \preprint coands are allowed.
% Use the 'preprintnumbers' class option to override journal defaults
% to display numbers if necessary
%\preprint{}
%Title of paper
\title{Magnetic phase diagram of Cu$_{4-x}$Zn$_x$(OH)$_6$FBr studied by neutron-diffraction and $\mu$SR techniques}
\author{Yuan Wei}
\affiliation{Beijing National Laboratory for Condensed Matter Physics, Institute of Physics, Chinese Academy of Sciences, Beijing 100190, China}
\affiliation{School of Physical Sciences, University of Chinese Academy of Sciences, Beijing 100190, China}
\author{Xiaoyan Ma}
\affiliation{Beijing National Laboratory for Condensed Matter Physics, Institute of Physics, Chinese Academy of Sciences, Beijing 100190, China}
\affiliation{School of Physical Sciences, University of Chinese Academy of Sciences, Beijing 100190, China}
\author{Zili Feng}
\affiliation{Beijing National Laboratory for Condensed Matter Physics, Institute of Physics, Chinese Academy of Sciences, Beijing 100190, China}
\affiliation{Institute for Solid State Physics, University of Tokyo, Kashiwa 277-8581, Japan}
\author{Devashibhai Adroja}
\affiliation{ISIS Facility, Rutherford Appleton Laboratory, Chilton, Didcot Oxon OX11 0QX, United Kingdom}
\affiliation{Highly Correlated Matter Research Group, Physics Department, University of Johannesburg, PO Box 524, Auckland Park 2006, South Africa}
\author{Adrian Hillier}
\affiliation{ISIS Facility, Rutherford Appleton Laboratory, Chilton, Didcot Oxon OX11 0QX, United Kingdom}
\author{Pabitra Biswas}
\affiliation{ISIS Facility, Rutherford Appleton Laboratory, Chilton, Didcot Oxon OX11 0QX, United Kingdom}
\author{Anatoliy Senyshyn}
\affiliation{Heinz Maier-Leibnitz Zentrum (MLZ), Technische Universit\"{a}t M\"{u}nchen, Garching D-85747, Germany}
\author{Chin-Wei Wang}
\affiliation{Synchrotron Radiation Research Center, Hsinchu 30076, Taiwan}
\author{Andreas Hoser}
\affiliation{Helmholtz-Zentrum Berlin f\"{u}r Materialien und Energie, D-14109 Berlin, Germany}
\author{Jia-Wei Mei}
\affiliation{Shenzhen Institute for Quantum Science and Engineering, and Department of Physics, Southern University of Science and Technology, Shenzhen 518055, China}
\author{Zi Yang Meng}
\affiliation{Beijing National Laboratory for Condensed Matter Physics, Institute of Physics, Chinese Academy of Sciences, Beijing 100190, China}
\affiliation{Department of Physics and HKU-UCAS Joint Institute of Theoretical and Computational Physics, The University of Hong Kong, Pokfulam Road, Hong Kong, China}
\affiliation{Songshan Lake Materials Laboratory , Dongguan, Guangdong 523808, China}
\author{Huiqian Luo}
\email{hqluo@iphy.ac.cn}
\affiliation{Beijing National Laboratory for Condensed Matter Physics, Institute of Physics, Chinese Academy of Sciences, Beijing 100190, China}
\affiliation{Songshan Lake Materials Laboratory , Dongguan, Guangdong 523808, China}
\author{Youguo Shi}
\email{ygshi@iphy.ac.cn}
\affiliation{Beijing National Laboratory for Condensed Matter Physics, Institute of Physics, Chinese Academy of Sciences, Beijing 100190, China}
\affiliation{Songshan Lake Materials Laboratory , Dongguan, Guangdong 523808, China}
\author{Shiliang Li}
\email{slli@iphy.ac.cn}
\affiliation{Beijing National Laboratory for Condensed Matter Physics, Institute of Physics, Chinese Academy of Sciences, Beijing 100190, China}
\affiliation{School of Physical Sciences, University of Chinese Academy of Sciences, Beijing 100190, China}
\affiliation{Songshan Lake Materials Laboratory , Dongguan, Guangdong 523808, China}
\begin{abstract}
We have systematically studied the magnetic properties of Cu$_{4-x}$Zn$_x$(OH)$_6$FBr by the neutron diffraction and muon spin rotation and relaxation ($\mu$SR) techniques. Neutron-diffraction measurements suggest that the long-range magnetic order and the orthorhombic nuclear structure in the $x$ = 0 sample can persist up to $x$ = 0.23 and 0.43, respectively.  The temperature dependence of the zero-field (ZF) $\mu$SR spectra provide two characteristic temperatures, $T_{A0}$ and $T_{\lambda}$. Comparison between $T_{A0}$ and $T_M$ from previously reported magnetic-susceptibility measurements suggest that the former comes from the short-range interlayer-spin clusters that persist up to $x$ = 0.82. On the other hand, the doping level where $T_{\lambda}$ becomes zero is about 0.66, which is much higher than threshold of the long-range order, i.e., $\sim$ 0.4. Our results suggest that the change in the nuclear structure may alter the spin dynamics of the kagome layers and a gapped quantum-spin-liquid state may exist above $x$ = 0.66 with the perfect kagome planes.
\end{abstract}

% insert suggested PACS numbers in braces on next line


%\maketitle must follow title, authors, abstract, \pacs, and \keywords
\maketitle

Two-dimensional (2D) kagome antiferromagnetic (AFM) system has attracted great interests since its strong geometrical frustration effects can give rise to various ground states \cite{MessioL12,BieriS15,IqbalY15,GongSS16,LiaoHJ17,MeiJW17,WangYC17}. Especially, it has been suggested to be one of the best platforms to realize quantum spin liquids (QSLs), which are highly entangled quantum magnetism that typically, although not necessarily, shows no long-range magnetic order down to zero K \cite{SavaryL17,ZhouY17}. Up to now, the most well-studied kagome $S$ = 1/2 magnetic material is the herbertsmithite, ZnCu$_3$(OH)$_6$Cl$_2$, which is believed to host a QSL ground state \cite{NormanMR16,ShoresMP05,BertF07,MendelsP07}. However, there is no consensus on whether its low energy excitations are gapped or gapless \cite{HeltonJS07,VriesMA09,HanTH12,HanTH16,FuM15,KhuntiaP20}, which is crucial for us to understand the nature of the QSL state.  One of the major difficulties lies in the fact that there are always a few percent of Cu$^{2+}$ ions sitting on Zn$^{2+}$ sites, which mainly affect the low-energy spin excitations \cite{HanTH16,NilsenGJ13}. 

Recently, there are increasing new materials that also consist of 2D Cu$^{2+}$ kagome layers \cite{SunW16,OkumaR17,FengZL17,FengZL18b,PuphalP18,WeiY19,IidaK20}. While most of them have magnetic orders, the Zn-doped barlowite (Cu$_3$Zn(OH)$_6$FBr) provides a promising new platform to study the QSL physics. The structure of barlowite Cu$_4$(OH)$_6$FBr is composed of 2D Cu$^{2+}$ kagome layers with Cu$^{2+}$ ions in between and shows long-range AFM order at 15 K\cite{HanTH14,FengZL18,TustainK18}. Substituting interlayer Cu$^{2+}$ with Zn$^{2+}$ can completely destroy the order and when the substitution is 100\%, one expects a QSL ground state \cite{FengZL18,SmahaRW20,TustainK20}.  Indeed, the system shows no magnetic ordering down to 50 mK although its dominate AFM exchange interaction is about 200 K \cite{FengZL17}. More interestingly, both NMR and inelastic neutron scattering results suggest  it has a gapped QSL ground state \cite{FengZL17,WeiY17}. Compared to the herbersmithite, it has higher crystal symmetry, which gives rise to smaller Dzyaloshinskii-Moriya interaction (DMI) \cite{HanTH16b,ZorkoA08} and thus makes it further away from the quantum phase transition resulted from the DMI interaction \cite{CepasO08}. Moreover, symmetry lowering has been observed in the herbertsmithite but not in Cu$_3$Zn(OH)$_6$FBr \cite{ZorkoA17,LauritaNJ19,NormanMR20,LiY20}. 

While the Zn-doped barlowite shows its advantages in studying the QSL physics on a 2D kagome lattice, it still suffers the same magnetic-impurity issues as the herbertsmithite \cite{FengZL17,WeiY17}. To further address the role of interlayer Cu$^{2+}$ spins, we systematically study the Cu$_{4-x}$Zn$_x$(OH)$_6$FBr system by the neutron-diffraction and $\mu$SR techniques. The long-range magnetic order and low-temperature orthorhombic structure are observed in the neutron-diffraction experiments for $x$ up to 0.23 and 0.43, respectively. The ZF $\mu$SR spectra can be fitted by a phenomenological function, which gives two parameters $A_0$ and $\lambda$, corresponding to the extrapolated zero-time asymmetry and long-time relaxation rate, respectively. By combining all the results, we provide a phase diagram of the Cu$_{4-x}$Zn$_x$(OH)$_6$FBr system by combining previous bulk and $\mu$SR results \cite{FengZL18,TustainK20}. Our results are consistent with a gapped ground state in Cu$_3$Zn(OH)$_6$FBr.

\begin{figure}[tbp]
\includegraphics[width=\columnwidth]{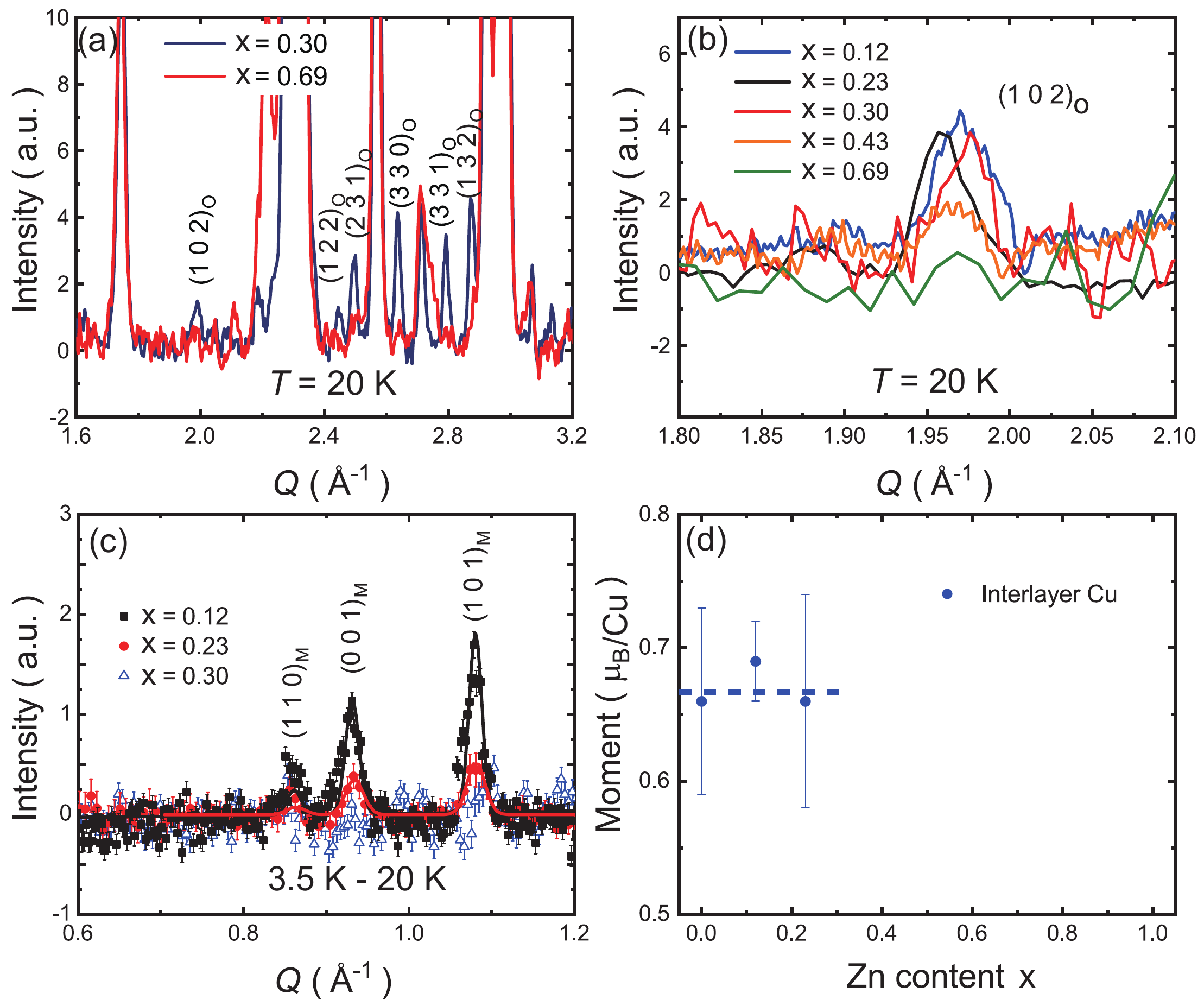}
 \caption{(a) Neutron powder diffraction intensities of the $x$ = 0.3 and 0.69 samples at 20 K in the range of large $Q$'s. The labeled peaks correspond to the orthorhombic peaks in the $Pnma$ space group. (b) The (1,0,2)$_O$ peaks for $x$ from 0.12 to 0.69 samples at 20 K. (c) First three magnetic peaks for the $x$ = 0.12, 0.23 and 0.3 samples obtained by subtracting the 20-K data from the 3.5-K data. The solid lines are calculated results for the magnetic structure as reported in \cite{FengZL18}. All the data in (a), (b) and (c) have been normalized by major nuclear peaks after background subtracted. It should be pointed out that since different instruments have different resolution, the normalization only provides a rough guide. (d) Temperature dependence of the magnetic moment for the interlayer Cu$^{2+}$ ions.}
\end{figure}

Polycrystalline Cu$_{4-x}$Zn$_x$(OH)$_6$FBr were synthesized by the hydrothermal method as reported previously \cite{FengZL18}. The neutron-diffraction data were obtained on SPODI at FRM-II, Germany, WOMBAT at ANSTO, Australia, and the instrument E9 at HZB, Germany. The nuclear and magnetic structures are refined by the FULLPROF program \cite{Rodriguez-CarvajalJ93}. The $\mu$SR experiments were carried out in the longitudinal field (LF) geometry at MuSR and EMU spectrometers of the ISIS Facility at the Rutherford Appleton Laboratory, Oxfordshire, U.K.  The samples were mounted on a 99.995\% silver plate, applying dilute GE varnish covered with a high-purity silver foil. The $\mu$SR data were analyzed by the MantidPlot software \cite{ArnoldO14}. 

Figure 1(a) shows the neutron-powder-diffraction intensities of the $x$ = 0.3 and 0.69 samples at 20 K. It has been shown that the low-temperature nuclear strucutre of the $x$ = 0 sample is orthorhombic with the space group of $Pnma$ \cite{FengZL18}. To see whether the orthorhombic structure exists in these two samples, the panel only plots the magnification of the data at large $Q$'s, which shows several orthorhombic peaks for the $x$ = 0.3 sample but not for the $x$ = 0.69 sample. Refinements on the data demonstrate that the lattice space group of the $x$ = 0.3 sample is $Pnma$, the same as that in the $x$ = 0 sample \cite{FengZL18}. The structure of the $x$ = 0.69 sample is hexagonal with the space group of $P6_3/mmc$ as that of the $x$ = 0.92 sample \cite{FengZL17}. Figure 1(b) gives the (1,0,2)$_O$ peak for different samples, which shows that the orthorhombic structure may still present in the $x$ = 0.43 sample. Future studies with large $Q$'s data are need to determine at which doping level the orthorhombic structure disappears.

\begin{figure}[tbp]
\includegraphics[width=\columnwidth]{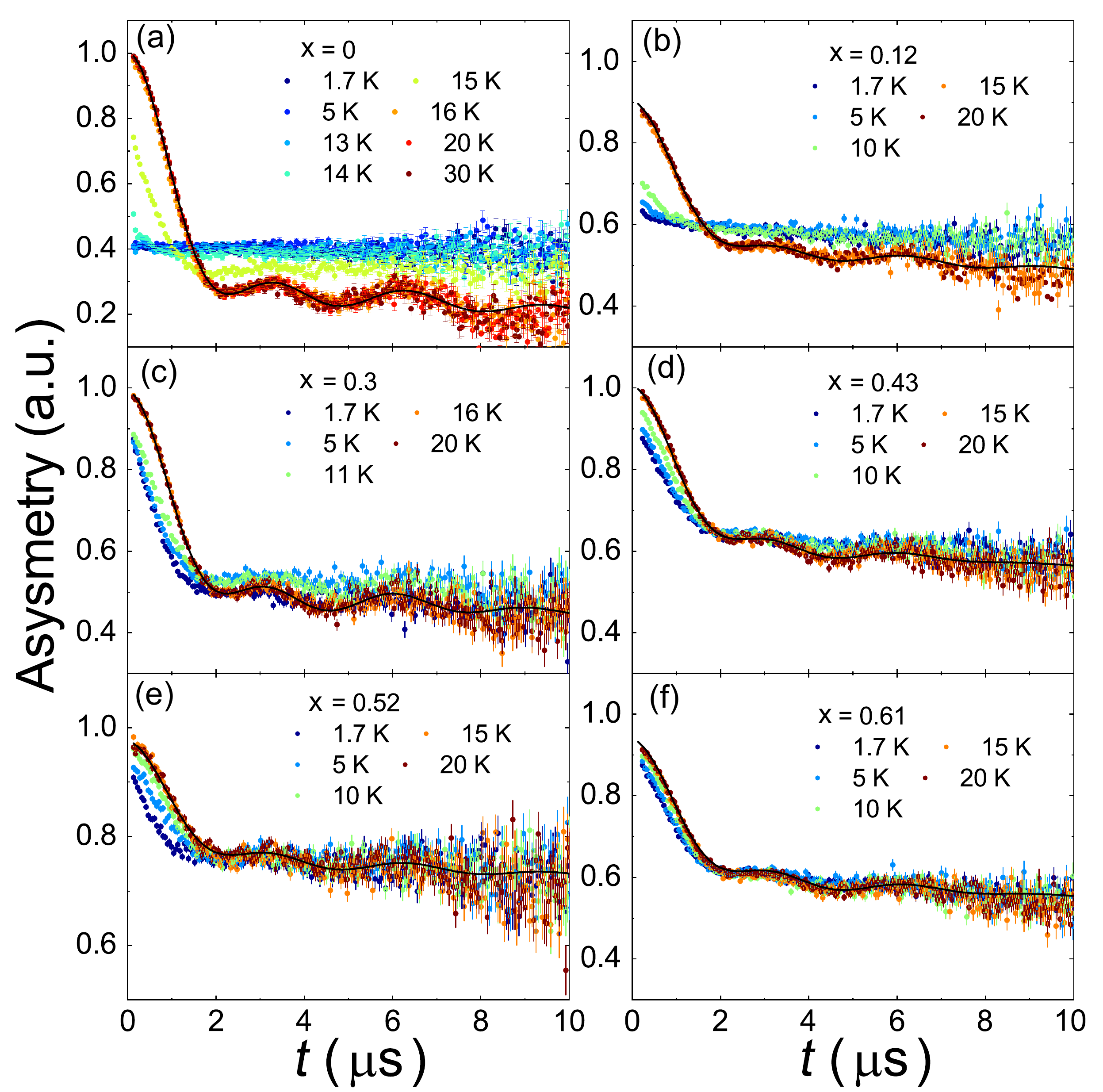}
 \caption{(a) - (f) Time-dependent ZF $\mu$SR spectra at different temperatures for the $x$ = 0, 0.12, 0.3, 0.43, 0.52 and 0.61 samples, respectively. The solid lines are fitted by Eq. (\ref{At}).}
\end{figure}

Figure 1(c) shows the first three magnetic peaks for the $x$ = 0.12 and 0.23 samples, which can be refined by the same magnetic structure as in barlowite \cite{FengZL18}. In this structure, the ordered moment mainly comes from the interlayer Cu$^{2+}$ spins, while the magnetic configuration for the kagome spins is rather hard to be determined due to their weak ordered moments \cite{FengZL18,TustainK18}. The doping dependence of the ordered interlayer moment is shown in Fig. 1(d), which suggests that its value does not change with doping. The decrease of the magnetic-peak intensities is mainly due to the substitution of nonmagnetic Zn$^{2+}$ ions. While our neutron-diffraction data cannot distinguish whether there is magnetic order for the $x$ = 0.3 sample because of its very weak signal (Fig. 1(c)), previous $\mu$SR has shown that the magnetically ordered phase can survive up to 0.32 \cite{TustainK20}.

\begin{figure}[tbp]
\includegraphics[width=\columnwidth]{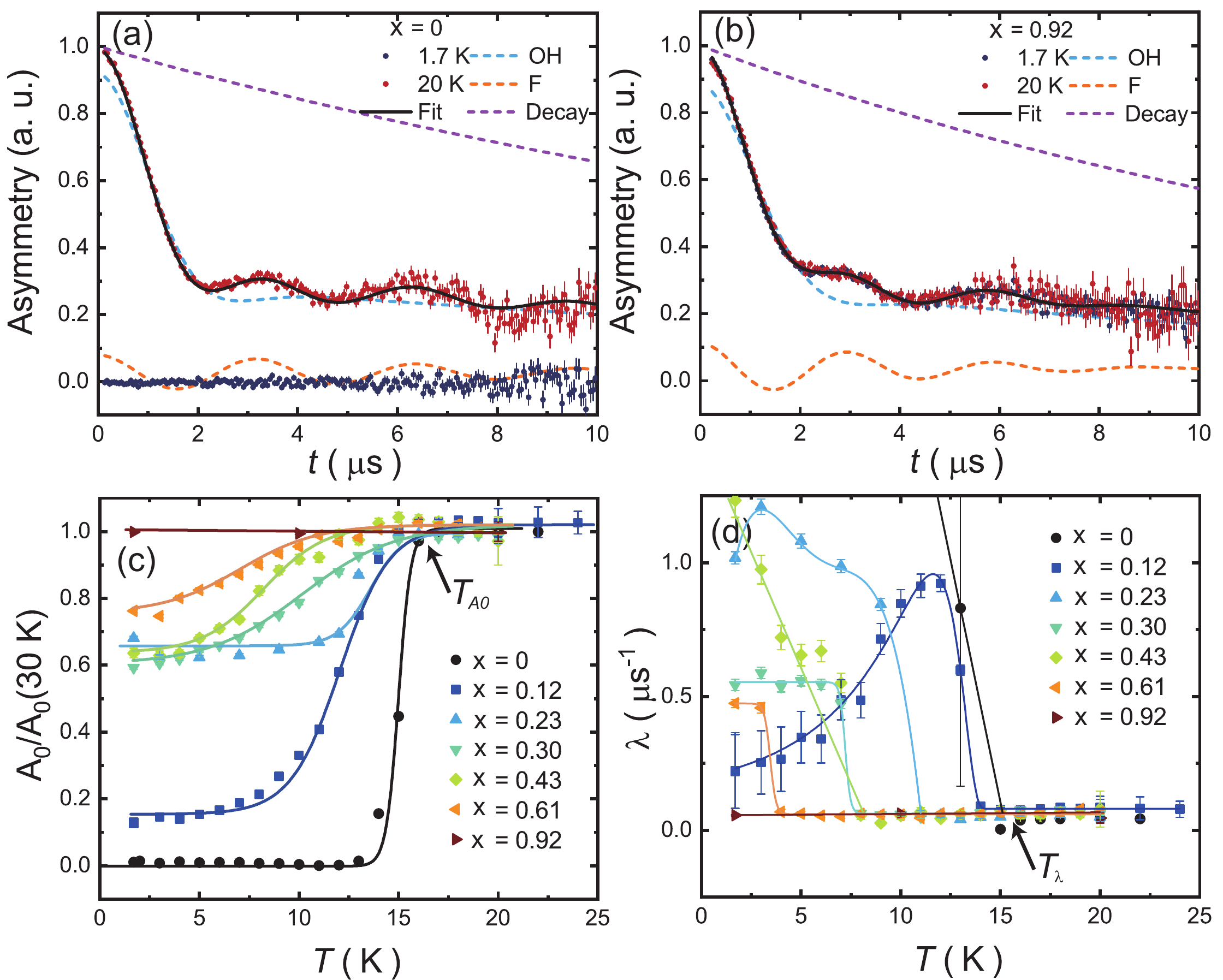}
 \caption{(a) \& (b) Detailed analysis of the $\mu$SR spectra for the $x$ = 0 and 0.92 samples, respectively. The solid lines are the fitted results according to Eq. (\ref{At}). The blue and red dashed lines are contributions from the $\mu$-OH and $\mu$-F complexes, respectively. The purple dashed lines show the contribution from the exponential decay. (c) \& (d) Temperature dependence of $A_0/A_0(30 K)$ and $\lambda$. The solid lines are guides to the eye. The arrows in (c) and (d) mark the transition temperatures of $A_0$ and $\lambda$ for the $x$ = 0 sample, respectively.}
\end{figure}

Figure 2 provides the time-dependent ZF $\mu$SR spectra for the Cu$_{4-x}$Zn$_x$(OH)$_6$FBr system. At high temperatures, all of them show oscillation behaviors. With decreasing temperature, the oscillation is completely suppressed in the time range measured here for the low-doping samples but still presents for the middle doping samples. It has been shown that the oscillation is associated with both $\mu$-OH and $\mu$-F complexes \cite{TustainK20}. The actual description of the $\mu$SR spectra needs detailed information of the asymmetry at very low time scale close to zero, which is not accessible for our data. We thus simply introduce the following equation to account for the oscillation,

\begin{equation}
D^i_z(t) = [\frac{1}{3}+\frac{2}{3}cos(\omega_it)]e^{-\sigma_i^2t^2},
\label{Dz}
\end{equation}
\noindent where $i$ denotes for the $\mu$-OH and $\mu$-F complexes for $i$ = 1 and 2, respectively, and $\sigma_i$ is associated with the distribution of nuclear fields surrounding the muon spin. The time dependence of the asymmetry can be written as follows,

\begin{equation}
A(t) = A_0[fD^1_z(t)+(1-f)D^2_z(t)]e^{-\lambda t} + A_{bg}
\label{At}
\end{equation}
\noindent where $\lambda$ describes the weak electronic relaxation of the muons stopping in the sample and $A_{bg}$ is a constant temperature-independent background for the muon stopping on the Ag sample holder. A similar function has also been used in studying  the Zn$_x$Cu$_{4-x}$(OH)$_6$Cl$_2$ system \cite{MendelsP07}. The factor $f$ is introduced to account for different contributions from $\mu$-OH and $\mu$-F complexes. For all the samples measured here, $f$ is found to be 0.88. It should be pointed out that although Eq. (\ref{At}) cannot precisely describe the actual muon relaxation, it provides good approximations of the $\mu$SR spectra for all the samples at all temperatures as shown in Fig. 2.

\begin{figure}[tbp]
\includegraphics[width=\columnwidth]{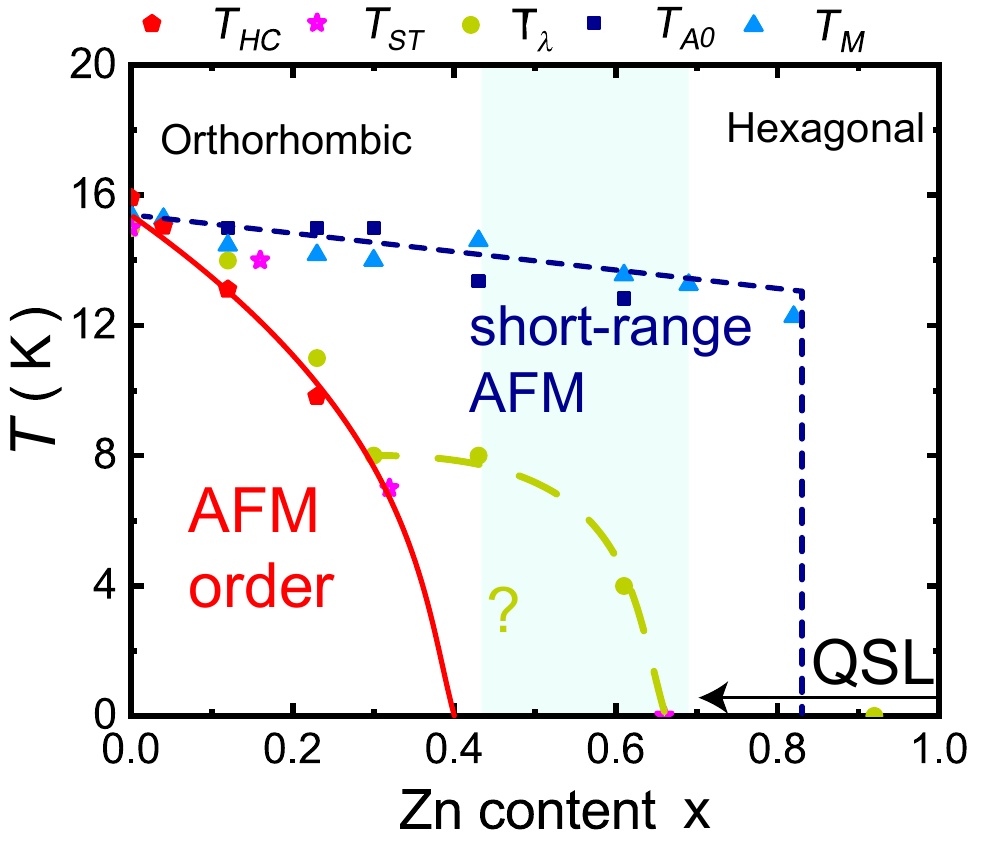}
 \caption{Phase diagram of Cu$_{4-x}$Zn$_x$(OH)$_6$FBr. The AFM order is the long-range 3D magnetic order involving both the intralayer and interlayer spins. The nuclear structures for the low-doping and high-doping samples are orthorhombic and hexagonal, respectively, with the crossover happens in the doping range marked by the shaded area. The short-range AFM is from the interlayer Cu$^{2+}$ ions. The question mark indicates the region between about 0.4 and 0.66 with an unknown ground state. The QSL ground state in the $x$ = 1 sample may be extended down to 0.66, as shown by the arrow. $T_{A0}$ and $T_\lambda$ are determined by the temperature dependence of $A_0$ and $\lambda$ as shown in Fig. 3(c) and 3(d). $T_{HC}$ and  $T_M$ are from the heat-capacity and magnetic-susceptibility measurements \cite{FengZL18}. 
$T_{ST}$ is from the temperature dependence of the frozen fraction from the $\mu$SR measurement reported previously \cite{TustainK20}. The dashed lines are guides to the eye.}
\end{figure}

Figure 3(a) and 3(b) further provide detailed analysis of the muon spectra for the $x$ = 0 and 0.92 samples, respectively. The initial fast drop of the asymmetry is mainly due to the $\mu$-OH complex, while the $\mu$-F complex mainly contributes to the oscillation above about 2 $\mu$s \cite{TustainK20}. The relaxation term exp$(-\lambda t)$ is also determined by the long-time data, which may be related to the $\mu$-F complex. Figure 3(c) and 3(d) show the temperature dependence of the fitted parameters $A_0$ normalized by its 30-K value and $\lambda$, respectively. For the $x$ = 0 sample, significant changes are found at $T_N$ for both $A_0$ and $\lambda$. The temperatures are marked as $T_{A0}$ and $T_{lambda}$, respectively. With increasing Zn substitution, both $T_{A0}$ and $T_{\lambda}$ decrease but with different rate. While the former is still above 10 K for the $x$ = 0.61 sample, the latter has already dropped to about 4 K. 

Figure 4 gives the phase diagram of Cu$_{4-x}$Zn$_x$(OH)$_6$FBr by summarizing our results and previous studies \cite{FengZL18,TustainK20}. It has been previously suggested that the magnetic transition temperatures detected by the heat capacity and magnetic susceptibility, i.e. $T_{HC}$ and $T_M$, become different in the Zn-substituted samples \cite{FengZL18}. Here we found that $T_{HC}$ is close to $T_{ST}$, which is defined as the temperature where the frozen volume fraction obtained from the $\mu$SR spectra becomes non-zero \cite{TustainK20}. Since both the heat capacity and frozen volume fraction are measuring the bulk properties, and our neutron-diffraction data also clearly show magnetic peaks up to $x$ = 0.23, we conclude that the region for the long-range order AFM order are marked by $T_{HC}$ and $T_{ST}$.

Figure 4 also demonstrates that $T_M$ and $T_{A0}$ are very close to each other, which suggests they have the same origin. As shown above, the value of $A_0$ is not the asymmetry at time zero but rather the extrapolated zero-time value after the very-fast initial drop that cannot be detected here. For the $x$ = 0 sample, this kind of drop is due to the local fields from the long-range magnetic order \cite{TustainK20}. With Zn substitution, the long-range order is suppressed but the magnetic clusters, most likely formed by the interlayer Cu$^{2+}$ ions, can still survive and provide local magnetic fields to depolarize the muon spins, which will result in the fast initial drop of the asymmetry. This is consistent with previous discussions on the origin of the $T_M$ \cite{FengZL18}. Therefore, the area marked by $T_M$ and $T_{A0}$ is associated with the short-range AFM from the interlayer moments. It should be noted that for low-doping samples, the system enters into the short-range AFM first and then becomes long-range ordered with decreasing temperature, which suggests that the long-range AFM order should involve kagome spins.

Tracing the doping dependence of $T_{\lambda}$ shows that it ends at about $x$ = 0.66, which is clearly not associated with either the long-range or the short-range AFM. While its origin is unclear, there are a few results that may be related to it. First, the crossover from the orthorhombic to the hexagonal structure may also happen at the same doping range, as illustrated by the shaded area in Fig. 4. The change of the structure should result in the change of the exchange energies within the kagome planes. Second, it has been shown that the $\mu$SR spectra do not change any more for $x \geq$ 0.66 \cite{TustainK20}, which suggests that the ground states for 0.66 $\leq x \leq$ 1 may be the same. The regime for 0.32 $< x <$ 0.66 is suggested to be the crossover between the static and dynamic ground states. This is consistent with our observation of the regime between red line for the AFM order and green dashed line for $T_{\lambda}$, where the ground state of kagome planes is changing from the 3D order to the gapped quantum spin liquid though an unknown intermediate state. Third, the value of $\lambda$ is mainly determined by the long-time relaxation and thus related to the $\mu$-F complex. The position of the fluorine makes it not sensitive to interlayer spins, as shown by the $^{19}$F NMR measurements \cite{FengZL17}. As the low-energy spin excitations become gapped \cite{FengZL17,WeiY17}, one expect that $\mu$SR spectra from the $\mu$-F complex, which only detect very low-energy excitations, will not change with the temperature any more. These results suggest that the change of $\lambda$ at low temperatures for $x <$ 0.66 may be related to the kagome spins and the gapped QSL state in the $x$ = 1 sample could be extended down to 0.66. It should be noted that the unknown crossover region labeled by the question mark and the QSL state should not be treated as within the short-range magnetic order, but rather as solely coming from the kagome spin system, which is phase-separated from the interlayer spin system.

In conclusions, we establish the magnetic phase diagram of the Cu$_{4-x}$Zn$_x$(OH)$_6$FBr system by comparing several techniques \cite{FengZL18,TustainK20}. The short-range spin correlations of the interlayer Cu$^{2+}$ moments can persist up to $x \approx$ 0.82, while the whole spin system only become long-range ordered below  $x \approx$ 0.4. The slow exponential decay of the long-time $\mu$SR spectra disappears above $x$ = 0.66, which is consistent with a gapped QSL state. Our results suggest that the kagome and interlayer spin systems are decoupled at high-Zn-doping levels and may help us to further understand the actual magnetic ground state of the 2D kagome antiferromagnet Cu$_3$Zn(OH)$_6$FBr.

\acknowledgments

This work is supported by the National Key Research and Development Program of China (Grants No. 2017YFA0302900, No. 2016YFA0300500, No. 2018YFA0704200, No. 2017YFA0303100, No. 2016YFA0300600), the National Natural Science Foundation of China (Grants No. 11874401, No. 11674406, No.11674372, No. 11961160699, No.11774399, No.12061130200, No.11974392, No. 11822411)), the Strategic Priority Research Program(B) of the Chinese Academy of Sciences (Grants No. XDB25000000, No. XDB07020000, No. XDB33000000 and No. XDB28000000), Beijing Natural Science Foundation (No. Z180008,  No. JQ19002), Guangdong Introducing Innovative and Entrepreneurial Teams (No. 2017ZT07C062). H.L. is grateful for support from the Youth Innovation Promotion Association of CAS (Grant No. 2016004). X. M. H. L. and D.A. acknowledge the funding support of the Royal Society-Newton Advanced Fellowship  (Grant No. 
 NAF$\backslash$R1$\backslash$201248 ).

Y. W. and X. M. contributed equally to this work.

\end{document}